\documentclass[twocolumn,showpacs,preprintnumbers,amsmath,amssymb]{revtex4}

\usepackage{graphicx}
\usepackage{dcolumn}
\usepackage{bm}

\begin{document} 

\preprint{}

\title{Epitaxial Growth of a Full-Heusler Alloy Co$_{2}$FeSi on Silicon by Low-Temperature Molecular Beam Epitaxy}

\author{S. Yamada,$^{1}$ K. Yamamoto,$^{1}$ K. Ueda,$^{1}$ Y. Ando,$^{1}$ K. Hamaya,$^{1,2}$\footnote{E-mail: hamaya@ed.kyushu-u.ac.jp} T. Sadoh,$^{1}$ and M. Miyao$^{1}$}
\affiliation{%
$^{1}$Department of Electronics, Kyushu University, 744 Motooka, Fukuoka 819-0395, Japan}%
\affiliation{%
$^{2}$PRESTO, Japan Science and Technology Agency, 4-1-8 Honcho, Kawaguchi 332-0012, Japan}%


%

\date{\today}
\begin{abstract}
For electrical spin injection and detection of spin-polarized electrons in silicon, we explore highly epitaxial growth of ferromagnetic full-Heusler-alloy Co$_{2}$FeSi thin films on silicon substrates using low-temperature molecular beam epitaxy (LTMBE). Although {\it in-situ} reflection high energy electron diffraction images clearly show two-dimensional epitaxial growth for growth temperatures ($T_\text{G}$) of 60, 130, and 200 $^{\circ}$C, cross-sectional transmission electron microscopy experiments reveal that there are single-crystal phases other than Heusler alloys near the interface between Co$_{2}$FeSi and Si for $T_\text{G} =$ 130 and 200 $^{\circ}$C. On the other hand, almost perfect heterointerfaces are achieved for $T_\text{G} =$ 60 $^{\circ}$C. These results and magnetic measurements indicate that highly epitaxial growth of Co$_{2}$FeSi thin films on Si is demonstrated only for $T_\text{G} =$ 60 $^{\circ}$C.
\end{abstract}

\maketitle
Lots of researchers have studied solutions to overcome the scaling limits of existing silicon-based complementary metal-oxide-semiconductor (CMOS) technologies. Hereafter, we focus on approaches using spin function of electrons to existing semiconductor technologies, i.e., semiconductor spintronics. This enables us to remarkably improve device performance such as nonvolatility, reconstrucutibility, low power consumption, and so on. For III-V semiconductor spintronic studies,\cite{Hanbicki} crystal growth of ferromagnetic epilayers on semiconductors is one of the key technologies to efficiently inject spin-polarized carriers from ferromagnets into semiconductors. Recently, we have studied epitaxial growth of ferromagnetic Heusler alloys on group-IV semiconductors such as Si or Ge by low-temperature molecular beam epitaxy (LT-MBE) and have realized high-quality heterostructures with atomically abrupt interfaces.\cite{Hamaya,Sadoh,Ueda} 
Expanding this LT-MBE technique for highly spin-polarized full-Heusler alloys is further promising to develop Si-based spintronics. As an ideal material compatible with Si, we focus on a ternary Heusler-alloy Co$_{2}$FeSi, which is theoretically expected to be half-metallic and a member of full-Heusler alloys with the fcc $L2_\text{1}$ crystal structure, where two Fe atoms of $DO_\text{3}$-type Fe$_{3}$Si are just substituted for two Co atoms.\cite{Niculescu} The bulk lattice constant ($\sim$ 0.566 nm) of Co$_{2}$FeSi nearly corresponds to that of Fe$_{3}$Si ($\sim$ 0.565 nm). Also, bulk Co$_{2}$FeSi has a large magnetic moment of ~5.49 $\mu_\text{B}$/f.u. at room temperature and exhibits ferromagnetic nature above 900 K,\cite{Niculescu} which is one of the highest Curie temperature among reported full-Heusler alloys. Recently, relatively high tunneling magnetoresistance (TMR) ratios\cite{Inomata} and highly efficient spin injection into (Al,Ga)As\cite{Ramsteiner} were demonstrated for spin devices with the Co$_{2}$FeSi electrodes.

In this paper, we demonstrate highly epitaxial growth of Co$_{2}$FeSi thin films on Si by LT-MBE. Cross-sectional transmission electron microscopy measurements reveal that the Co$_{2}$FeSi/Si heterointerface has almost no interlayer only for $T_\text{G} =$ 60 $^{\circ}$C. $L2_\text{1}$ ordered structures and magnetic properties, similar to that of the optimized Fe$_{3}$Si/Si, are clearly obtained.
\begin{figure}[b]
\begin{center}
\includegraphics[width=7.5cm]{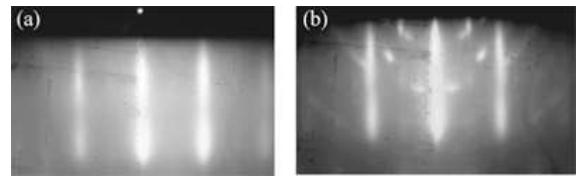}
\caption{Representative reflection high energy electron diffraction (RHEED) patterns observed along the [$\bar{2}$11] azimuth for Co$_{2}$FeSi surfaces at $T_\text{G} =$ (a) 60 $^{\circ}$C and (b) 400 $^{\circ}$C.}
\end{center}
\end{figure}    

Prior to the growth of Co$_{2}$FeSi thin films, we performed the substrate cleaning with an aqueous HF solution (HF : H$_{2}$O = 1 : 40) and a heat treatment at 450 $^{\circ}$C for 20 min with a base pressure of 2 $\times$ 10$^{-9}$ Torr. After a reduction in the substrate temperature down to $T_\text{G} =$ 60, 130, 200, 300, or 400 $^{\circ}$C, Co$_{2}$FeSi thin films were grown directly on the cleaned Si(111) surface by LT-MBE.\cite{Sadoh,Ueda,Maeda} Here, we co-evaporated Co, Fe, and Si with growth rates of 1.32, 0.72, and 1.20 nm/min, respectively, using Knudsen cells.
\begin{figure}[t]
\begin{center}
\includegraphics[width=5.5cm]{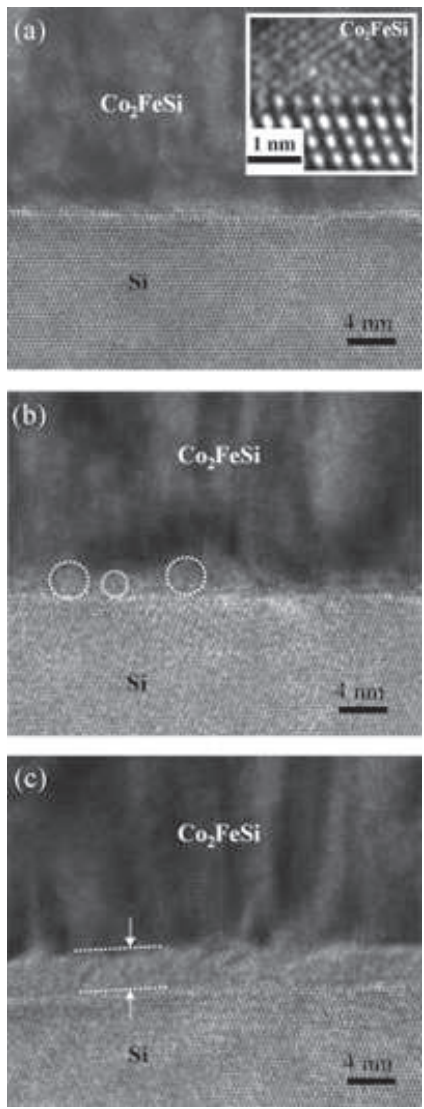}
\caption{Cross-sectional TEM images of the Co$_{2}$FeSi/Si(111) interfaces fabricated by LT-MBE for (a) $T_\text{G} =$ 60, (b) 130, and (c) 200 $^{\circ}$C. The inset of (a) is a high-resolution TEM image.}
\end{center}
\end{figure}  

During the growth, the observation of {\it in-situ} reflection high energy electron diffraction (RHEED) patterns was performed for all the samples. Representative RHEED patterns of the Co$_{2}$FeSi layers grown at $T_\text{G} =$ 60 and 400 $^{\circ}$C are shown in Figs. 1(a) and 1(b), respectively. In Fig. 1(a) the RHEED image clearly exhibits symmetrical streaks, implying single crystal and good two-dimensional epitaxial growth of the Co$_{2}$FeSi layers on Si(111). For $T_\text{G} \le$ 200 $^{\circ}$C, almost the same patterns were observed. In Fig. 1(b), on the other hand, we can see spotty and ring-like RHEED patterns, indicating three-dimensional and polycrystalline growth, together with light streaks. For $T_\text{G} \ge$ 300 $^{\circ}$C, similar features were observed. Hence, we firstly choose $T_\text{G}$ below 200 $^{\circ}$C. After the growth, x-ray diffraction (XRD) measurements of the Co$_{2}$FeSi layers were also carried out in $\theta-$2$\theta$ and 2$\theta$ configurations for $T_\text{G} =$ 60, 130, and 200 $^{\circ}$C. Since the lattice mismatch between Co$_{2}$FeSi and Si is about 4 \% for bulk samples, the diffraction peaks can be distinguished in $\theta-$2$\theta$ configuration. Although the intensity of the diffraction peak due to Co$_{2}$FeSi(111) was quite weaker than that due to Si(111) and we could not see it, which was also discussed for Fe$_{3}$Si/Si(111),\cite{Hamaya} we could observe the peak due to Co$_{2}$FeSi(222). We also confirmed no peak due to other compounds in 2$\theta$ configuration. Considering these results, we deduced that epitaxial growth of Co$_{2}$FeSi layers on Si(111) is achieved with no secondary phase for $T_\text{G} =$ 60, 130, and 200 $^{\circ}$C. 

To evaluate the Co$_{2}$FeSi/Si interface which will become important to inject and detect electron's spins, we measured cross-sectional transmission electron micrograph (TEM). Figures 2 (a)-(c) display the TEM images for $T_\text{G} =$ 60, 130, and 200 $^{\circ}$C, respectively. First, for $T_\text{G} =$ 200 $^{\circ}$C [Fig. 2(c)], we clearly identify $\sim$ 4-nm-thick interlayers at the interface between Co$_{2}$FeSi and Si (see-arrows). Nanobeam electron diffraction (nano-ED) measurements in this region detected some cubic single-crystal phases, which may be CsCl- and/or CaF$_{2}$-type silicides such as nonmagnetic CoSi, FeSi, and/or CoSi$_{2}$.\cite{Hong} These facts could not be identified by the above XRD measurements. Though we also measured nanobeam energy dispersive x-ray spectroscopy (nano-EDX) of the interlayers, we could not determine the materials because of the large fluctuations of the atomic compositions of Co, Fe, and Si. In Fig. 2(b) ($T_\text{G} =$ 130 $^{\circ}$C), we can recognize that the interlayer materials shown in Fig. 2(c) are reduced due to a decrease in $T_\text{G}$, but they can be detected yet, as indicated in the white dotted circles. Also, nano-ED patterns evidenced the presence of the above cubic single-crystal silicides. On the other hand, for $T_\text{G} =$ 60 $^{\circ}$C [Fig. 2(a)], we can evidently see a flat interface within the fluctuation of a few monolayers. The high-resolution image shown in the inset indicates nearly perfect interface between Co$_{2}$FeSi and Si. By nano-ED, we confirmed that there is no interlayer material described in Figs. 2(b) and 2(c). Thus, it is necessary for obtaining high-quality Co$_{2}$FeSi/Si interfaces to reduce $T_\text{G}$ down to 60 $^{\circ}$C, where this $T_\text{G}$ is fairly lower than that for obtaining high-quality Fe$_{3}$Si/Si interfaces ($T_\text{G} =$ 130 $^{\circ}$C).\cite{Hamaya} 

We concentrate on the high-quality epitaxial Co$_{2}$FeSi/Si grown at $T_\text{G} =$ 60 $^{\circ}$C. Depth profile of the atomic composition measured by nano-EDX revealed the achievement of almost stoichiometric Co$_{2}$FeSi. A nano-ED image of the Co$_{2}$FeSi layer is shown in the inset of Fig. 3(a). We find superlattice reflections, ($\overline{1}$$\overline{1}$1) and ($\overline{3}$$\overline{1}$1), which are caused by the presence of an ordered $L2_\text{1}$ structure (red dotted circles), together with superlattice reflections of the ordered $B2$ + $L2_\text{1}$ structures (blue dotted circles). Namely, even for $T_\text{G} =$ 60 $^{\circ}$C, good ordered single-crystal structures can be fabricated by the LT-MBE technique.
\begin{figure}[t]
\begin{center}
\includegraphics[width=7.0cm]{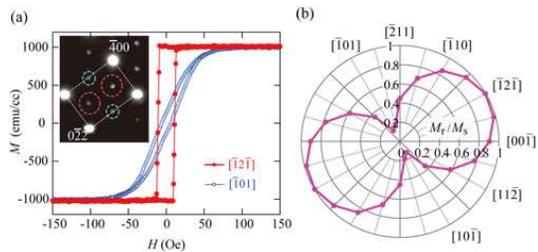}
\caption{(Color online) (a) Representative magnetization curves of the Co$_{2}$FeSi/Si(111) epilayer grown at $T_\text{G} =$ 60 $^{\circ}$C, measured along the magnetic easy ([$\overline{1}$2$\overline{1}$]) and hard ([$\overline{1}$01]) axes in the film plane at room temperature. The inset shows a nano-ED image of the 60 $^{\circ}$C-grown layer. (b) A polar plot of $M_\text{r}$/$M_\text{S}$.}
\end{center}
\end{figure} 

In order to examine magnetic properties of the Co$_{2}$FeSi/Si(111) epilayer, we also measured the field-dependent magnetization, $M-H$ curves, by vibrating sample magnetometer (VSM) at room temperature, where the applied filed was parallel to the layer plane. When external magnetic fields are applied parallel to crystallographic orientations of [$\overline{1}$2$\overline{1}$] and [$\overline{1}$01], we obtain the red and blue $M-H$ curves, respectively, as shown in Fig. 3(a). These data yield the saturation magnetization ($M_\text{S}$) of $\sim$ 1000 emu/cm$^{3}$ ($\sim$ 4.90 $\mu_\text{B}$/f.u.), close to that of bulk Co$_{2}$FeSi ($\sim$ 1124 emu/cm$^{3}$, $\sim$ 5.49 $\mu_\text{B}$/f.u.).\cite{Niculescu} We note that there is a strong in-plane magnetic anisotropy. To further investigate the anisotropic features, we also measured the $M-H$ curves for various crystallographic axes. Shown in Fig. 3(b) is summary of the remanent magnetization ($M_\text{r}$) normalized by $M_\text{S}$, i.e., $M_\text{r}$/$M_\text{S}$. Surprisingly, an unexpected uniaxial magnetic anisotropy with a magnetic easy axis along [$\overline{1}$2$\overline{1}$] can be detected. This uniaxial anisotropy cannot be explained only by the magnetocrystalline anisotropy of the single-crystal Co$_{2}$FeSi/Si(111), since it has six-fold crystal symmetry in the film plane. In our previous works, we have already observed a similar in-plane uniaxial anisotropy for Fe$_{3}$Si/Ge(111) and Fe$_{3}$Si/Si(111) epilayers with high-quality heterointerfaces,\cite{Ando} however, its precise origin is still unclear. For both cases, once the interlayers such as FeGe and FeSi were observed at the heterointerfaces, the above uniaxial anisotropy vanished.\cite{Ando} For the present Co$_{2}$FeSi/Si(111) epilayers, the observed uniaxial anisotropy shown in Fig. 3(b) was also reduced gradually with increasing $T_\text{G}$ and completely disappeared for $T_\text{G} =$ 300  and 400 $^{\circ}$C, similar to the features which were observed for the Fe$_{3}$Si/Ge(111) and Fe$_{3}$Si/Si(111) epilayers.\cite{Ando} These facts also support that the presence of the in-plane uniaxial anisotropy is probably caused by the high-quality Co$_{2}$FeSi/Si(111) interfaces. 

In summary, for silicon-based spintronics, we have explored the epitaxial growth of full-Heusler alloy Co$_{2}$FeSi thin films on Si at various $T_\text{G}$ by LT-MBE. For achieving high-quality heterointerfaces between Co$_{2}$FeSi  and Si, we have to reduce $T_\text{G}$ down to 60 $^{\circ}$C, which is lower than that for high-quality Fe$_{3}$Si/Si interfaces previously shown. 

The authors thank Prof. Y. Maeda of Kyoto University for useful discussion and experimental supports. This work was partly supported by a Grant-in-Aid for Scientific Research on Priority Area (No.18063018) from the Ministry of Education, Culture, Sports, Science, and Technology in Japan, and and PRESTO, Japan Science and Technology Agency, and Iketani Science and Technology Foundation.

\end{document}